\documentclass{emulateapj}

\usepackage{natbib}
\usepackage{graphicx,subfigure}
\usepackage{amsmath}

\newcommand\lsim{\mathrel{\rlap{\lower4pt\hbox{\hskip1pt$\sim$}}
        \raise1pt\hbox{$<$}}}
\newcommand\gsim{\mathrel{\rlap{\lower4pt\hbox{\hskip1pt$\sim$}}
        \raise1pt\hbox{$>$}}}

\newcommand \mum{ \mu{\rm m} }
\newcommand \charsig{ \overset{\sim}{\sigma} }

\begin{document}

\title{Cosmological X-ray Scattering from Intergalactic Dust}
\author{Lia Corrales}
\author{Frits Paerels}
\affil{Columbia Astrophysics Laboratory and Department of Astronomy, Columbia University, 550 West 120th Street, New York, NY 10027}
\email{lia@astro.columbia.edu}

\begin{abstract}

High resolution X-ray imaging offers a unique opportunity to probe the nature of dust in the $z \lsim 2$ universe.  Dust grains $0.1 -  1 \ \mum$ in size will scatter soft X-rays, producing a diffuse ``halo'' image around an X-ray point source, with a brightness $\sim$ few \% confined to an arcminute-sized region.  We derive the formulae for scattering in a cosmological context and calculate the surface brightness of the scattering halo due to (i) an IGM uniformly enriched ($\Omega_{\rm  d} \sim 10^{-5}$) by a power-law distribution of grain sizes, and (ii) a DLA-type ($N_{\rm H} \sim 10^{21}$ cm$^{-2}$) dust screen at cosmological distances.  The morphology of the surface brightness profile can distinguish between the two scenarios above, place size constraints on dusty clumps, and constrain the homogeneity of the IGM. Thus X-ray scattering can gauge the relative contribution of the first stars, dwarf galaxies, and galactic outflows to the cosmic metallicity budget and cosmic history of dust.  We show that, because the amount of intergalactic scattering is overestimated for photon energies $<$ 1 keV, the non-detection of an X-ray scattering halo by Petric et al. (2006) is consistent with `grey' intergalactic dust grains ($\Omega_d \sim 10^{-5}$) when the data is restricted to the 1-8 keV band.  We also calculate the systematic offset in magnitude, $\delta m \sim 0.01$, for such a population of graphite grains, which would affect the type of supernova survey ideal for measuring dark energy parameters within $\sim 1 \%$ precision.

\end{abstract}



\section{Introduction}
\label{sec:Intro}

The existence of a population of intergalactic dust grains would have wide-reaching implications on our understanding of the universe.  In particular a ``grey'' population of dust grains, which extinct uniformly across optical wavelengths, would affect extragalactic surveys that require precise optical and infrared photometry.  X-ray astronomy provides a unique opportunity to detect and  characterize dust that may be missed by traditional detection methods, which consider background object colors \citep[e.g.][]{Wright1981,Men2010}.  The scattering cross-section of dust to X-ray light is highly sensitive to the grain size (of radius $a$), with $\sigma_{\rm sca} \propto a^4 E^{-2}$, and occurs over small angles $\lsim 1 ' \ (a/\mum)^{-1} \ (E/{\rm keV})^{-1}$ \citep[][]{MG1986, SD1998}.  The scattering effect produces a diffuse ``halo'' image around a point-source \citep[e.g.][]{Over1965,Martin1970}, potentially capable of being observed with high-resolution optics such as those on ${\it Chandra}$.  X-ray scattering halos have been observed around bright point-sources and gamma-ray burst afterglows that occur behind dusty clouds in the Milky Way interstellar medium (ISM) \citep[e.g.][]{Rolf1983,MG1986,Witt2001,Smith2006,Vau2006,Tiengo2010}.  By looking at extragalactic X-ray point sources (quasars) well above the Galactic plane, observers can characterize the quantity of dust in the intergalactic medium (IGM) \citep{Evans1985}.  We will derive the formulae for scattering in a cosmological context and discuss interesting topics for which X-ray scattering can yield insight into the high-$z$ universe.

Stars create elements heavier than helium and thereby are the main contributors of dust building-blocks.  The mass of intergalactic dust  thus depends on the cosmic history of star formation and dust dispersal mechanisms.  
\citet{Ag1999b} estimated that about $50\%$ of metals at $z \sim 0.5$ are present in the IGM, and if $\sim 50\%$ of those metals are locked up in dust, then $\Omega_{\rm dust}^{\rm IGM} \sim 10^{-5}$.  More recent determination of cosmological parameters would only change these estimates $\sim 25\%$.
Constraints on the amount of cosmic dust using the thermal history of the IGM agreed that $\Omega_{\rm dust}^{\rm IGM} (z=3) \lsim 10^{-5}$ for grain sizes $a \geq 0.01 \ \mum$ \citep{IK2003}.  
\citet{DL2009} put an upper limit on $\Omega_{\rm dust}^{\rm IGM}$ from $7 \times 10^{-5}$ to $1.5 \times 10^{-4}$, under the hypothesis that 10\% of the soft X-ray background is the result of dust scattered light from AGN.  From this assumption they also showed that the optical depth to X-rays must be $\lsim 0.15$ at $z = 1.5$.
A previous search for a cosmological scattering halo around QSO 1508+5714 at $z=4.3$ by \citet{Petric2006} yielded a null result.  They used this to place the constraint $\Omega_{\rm dust}^{\rm IGM} < 2 \times 10^{-6}$.  We will show that our models are consistent with their observation and that the limits placed on the amount of intergalactic dust can be relaxed (\S\ref{sec:QSO1508}).

In Section~\ref{sec:DustSources} we examine the potential enrichment sources and the relative amounts of dust we expect to find in the IGM.  The first stars in the early universe (Pop III) and feedback from early star forming galaxies would have enriched the IGM with metals, creating a fairly uniform dust distribution that is comoving with cosmological expansion.  We review the dust scattering cross-section for X-ray light in Section~\ref{sec:Cross-Section} and use it to discuss the case of scattering through a constant comoving number density of dust grains in Section~\ref{sec:UniformIGM}.  Quasar absorption systems, caused by dense regions of neutral hydrogen -- often with traces of ionized metals -- already contain signatures of dust extinction \citep{Men2005b,York2006}.  Of individual dust sources, Damped Lyman-$\alpha$ systems (DLAs) are the densest absorbers and thus have the highest probability of producing an X-ray scattering signature.  We evaluate this possibility in Section~\ref{sec:ClumpyIGM} by calculating the X-ray scattering profile from an infinitely large screen of dust with column densities typical of DLAs.

Due to the large distances associated with cosmological scattering, the angular extent of an X-ray scattered halo can put limits on the uniformity of the IGM or the size of a dusty cloud.  Due to the nature of small angle scattering, the halo image is sensitive to dust at redshifts $z \lsim 2$ and is relatively insensitive to the redshift of the background point source.  We discuss the distances and timescales associated with cosmological X-ray scattering in Section~\ref{sec:Zsensitivity}.  Observing a scattering halo, in addition to finding extragalactic reservoirs of dust, can provide information about the nature of star formation and galactic structure at high-$z$.  In Section~\ref{sec:QSO1508} we reevaluate QSO 1508+4714, used by \citet{Petric2006}, in the 1-8 keV band to show that large dust grains with $\Omega_{\rm dust}^{\rm IGM} = 10^{-5}$ cannot be ruled out.  We summarize the variety of implications in Section~\ref{sec:Discussion}.


\section{Potential Sources of Intergalactic Dust}
\label{sec:DustSources}

Intergalactic dust may be distributed uniformly as a result of star formation in the early universe.  If that is the case, an azimuthally symmetric X-ray scattering halo may be observed around a background point-source at a distance large enough to provide sufficient column density for scattering.  Dust may also be efficiently distributed into galactic halos or the IGM through pressure-driven feedback from galaxies.  This effect may be particularly important at $z \sim 2-3$, during the epoch of star formation, and for small galactic halos from which outflows can more easily escape.  Previous work has concluded that dust grains $> 0.1 \ \mum$ are more efficiently expelled than smaller grains because they are grey to optical light, hence receiving more radiation pressure, and are less susceptible to deceleration by gas drag \citep{Dav1998,Ferr1991}.  Finally, if very little dust ($\Omega_{\rm dust}^{\rm IGM} \lsim 10^{-6}$) is evenly distributed throughout the cosmos today, we could potentially observe X-ray scattering from objects that are already known to have high column densities of gas and metals.  Quasar absorption systems offer a convenient opportunity to observe X-ray scattering from dust that may be present in these gas reservoirs.

\subsection{The Earliest Star Formation}
\label{sec:PopIII}

Population III stars formed from the near-pristine primordial hydrogen gas originating from the Big Bang, are typically $\sim 40 - 100 M_{\odot}$, and form in dark matter halos $\sim 10^6 M_\odot$ between $z \sim 20-30$ \citep{Bromm2004,Hoso2011}.  These stars are very large because cooling below $10^3$ K can only proceed by collisional excitation of the rotational modes of H$_2$ molecules.  Fragmentation into smaller clumps is possible once gas becomes enriched to a particular threshold, above which the cooling rate from metals dominates over H$_2$ cooling and exceeds the heating rate due to gravitational collapse: [C/H] $\approx -3.5$ and [O/H] $\approx -3.1$ \citep{Bromm2004}.  Many consider this critical metallicity to mark the transition between the era of Pop III and Pop II stars, which have low metallicity but are able to form smaller ($\sim 1 \ M_\odot$) stars that are observed today in globular clusters, dwarf spheroidal galaxies, and galactic halos.  The large characteristic mass-scale of Pop III stars indicates that the majority of stars ended their lives as supernovae, distributing metals.  If some of these metals are locked up in dust, it would contribute to cooling and allow gas to fragment at metallicities well below the critical threshold for metal line cooling \citep{Schn2006}.  The recent observation of a low-mass, extremely metal poor star with [Z/H] $\sim -5$ \citep{Caffau2011} supports the hypothesis that the first generation of stars produced dust.

Because Pop III stars are massive and form in small halos, feedback from radiation and thermal injection from supernovae are significant.  Thus the first generation of stars can efficiently remove gas from mini-halos and mix metals into the IGM \citep{HW2002, Bromm2003, Abel2007, Wise2008}.
In simulations by \citet{Wise2008}, the second generation of Pop~III stars had metallicities $-5 <$~[Z/H]~$< -3$ and formed around $z = 20$;  the third generation reached metallicities [Z/H]~$\gsim -3$.  These serve as an upper limit, because the simulations assumed that every Pop~III star ended in a pair-instability supernova (PISN).  
At solar metallicity, the mass ratio of metals versus H and He is $\approx 0.02$ \citep{DraineBook}.  Using this value to scale the metal mass by metallicity, the total dust-to-gas mass ratio of the Pop III enriched IGM is
\begin{equation}
	\label{eq:RatioIII}
	\left( \frac{M_d}{M_{\rm H}} \right)_{\rm III} = 0.02 \ f_{\rm dep} \ \left( \frac{Z}{Z_\odot} \right)
\end{equation}
where $f_{\rm dep}$ is the mass fraction of metals locked in dust.  Dust models suggest that $f_{\rm dep}$ can be rather high for both PISNe (0.3-0.7) and Type-II SNe (0.2-1) \citep{Schn2004,TF2001}.  Taking the metallicity threshold above which hydrogen clouds fragment, we can make a reasonable guess that the first stars may have enriched the IGM to a uniform background level of [Z/H]$ \sim -4$.  Equation~\ref{eq:RatioIII} then implies $(M_d/M_{\rm H})_{\rm III} \leq 2 \times 10^{-6}$.

Compare this to the ratio implied by assuming a particular value for the dust mass density in the IGM:
\begin{equation}
	\label{eq:ratioIGM}
	\left( \frac{M_d}{M_{\rm H}} \right)_{\rm IGM} \approx 7 \times 10^{-6} \ \Omega_{\rm dust}^{\rm IGM} \ \bar{n}_{\rm H}^{-1} \ h_{75}^2
\end{equation}
where $h_{75}$ is $H_0$ in units of 75 km/s/Mpc, and $\bar{n}_{\rm H} \sim 10^{-7}$~cm$^{-3}$ is the average baryon density of the universe today.  If $\Omega_{\rm dust}^{\rm IGM} \sim 10^{-5}$, the first generation of stars would have contributed $\sim$1/350 of that dust.  If Pop III stars were responsible for the entirety of IGM dust, they would have to pollute the universe to [Z/H]$\sim -1.5$, and all of those metals must be locked up in dust.  This is not likely to occur before the onset of smaller stellar populations and galaxy evolution.  Feedback from the second generation of stars and early galaxies is therefore required to bring the overall dust density of the IGM to a threshold detectable via X-ray scattering.

\subsection{Feedback from Galaxies at High and Low-$z$}
\label{sec:Feedback}

There is direct evidence that L* galaxies today can expel dust out to a few kpc at least.  Polarized light, a consequence of scattering, was observed around the disk of the starburst galaxy M82 well before its dusty outflows were imaged by {\it Hubble} and {\it Spitzer} \citep{Schm1976}.  The outflows are fueled by starburst activity and can host super-solar abundances, a result of supernovae enrichment, effectively contaminating the IGM with metals \citep{Kon2011}.  

In theory, dust grains can be efficiently expelled from a galactic disk when the forces of radiation pressure, gravity, and gas drag are weighed.   Large grains ($\geq 0.1 \ \mum$) can be expelled completely from optically thin disks, while small grains are either retained close to their formation sites or trapped in layers within the galactic halo \citep{Dav1998,Gb1987}.  The expulsion is also more efficient for graphite grains $\sim 0.1 \ \mum$, which absorb more light across the visual spectrum than silicate grains of comparable size, and can reach velocities of several hundred km/s.  The smaller timescale for exiting the halo allows graphite grains $> 0.1 \ \mum$ to survive sputtering processes compared to silicate grains, which exit slowly and are thus more likely to be destroyed \citep{Bar1989,Ferr1991}.  On the other hand, graphite grains can accumulate more charge than silicate grains as they exit a galactic disk.  Coulomb forces prevent graphite grains from traveling as far as silicate grains with the same starting velocity, an effect that may allow silicate grains to populate a galactic halo or the IGM \citep{BF2005}.  Regardless, enrichment processes suggest that an IGM dust population is likely to contain grains $\gsim 0.1 \ \mum$ in size and is consequently grey to optical light.

Dust grain expulsion is not just a side-effect, but may be {\it necessary} for galactic outflows to occur.  \citet{MQT2005} argued that the Faber-Jackson relation between an elliptical galaxy's luminosity and velocity dispersion can be explained by the balance between momentum driven winds and stellar luminosity.  They also showed that, for radiation pressure to act efficiently, dust is required to contribute to the total ISM opacity.  As material is expelled, the gas may become sufficiently diffuse for the dust to decouple and migrate outwards alone.  They estimated the radius at which this occurs ($R_{dg}$), with respect to the velocity dispersion ($\sigma$), gas fraction ($f_g$), and dust grain size ($a$):
\begin{equation}
	R_{dg} \sim 150 \ a_{(0.1 \ \mum)} \ \sigma_{(200 \ {\rm km/s})}^2 \ f_{g,0.1} \ {\rm kpc}
\end{equation}
where $0.1 \ \mum$, 200 km/s, and 0.1 are fiducial values for $a$, $\sigma$, and $f_g$ respectively.
Dust grains $\sim 0.1 \ \mum$ expelled from very small halo potentials ($\sigma \sim 50$ km/s) would decouple from the gas $10$ kpc beyond the site of origin.  Thus during the epoch of star formation, gas may not reliably trace the spatial distribution of intergalactic dust.  If that is the case, the fraction of IGM metals locked up in dust may be rather large ($\sim 0.5$).

An estimate for the total metallicity of the universe as a function of redshift can be obtained by integrating over the observed star formations rates, yielding $\Omega_Z = 2.5 \times 10^{-5}$ at $z=2.5$ \citep{Pett2006}.  The total mass of metals observed in Ly-$\alpha$ absorption systems and Lyman Break Galaxies (LBGs) at $z = 2.5$ differs by the amount predicted by a factor of few, coined the ``missing metals problem'' \citep{Pagel2002,Pett2004}.  The discrepancy may be due to uncertainty in the star formation rates between $z = 4 - 10$ or galaxies that have been missed by UV-selection \citep{Pett2006}.  \citet{Bou2007} used analytic calculations to conclude that around 50\% of the total expected metal mass at $z=2$ could have been ejected by sub-L* (the majority of which comes from L $< 1/3$ L*) galaxies.  
If 50\% of intergalactic metals are locked up in dust, then $\Omega_{\rm dust}^{\rm IGM}(z \leq 2) \sim 8 \times 10^{-6}$ due to pollution by sub-L* galaxies during the epoch of star formation.  X-ray scattering halos can help test whether or not metals really have been missing from observations of the $z=2$ universe, and whether feedback mechanisms can account for the supposed deficiency.

\subsection{Damped Lyman-$\alpha$ Absorbers}
\label{sec:DLAs}

DLAs are dense regions of neutral hydrogen ($N_{\rm HI} \geq 2 \times 10^{20}$ cm$^{-2}$) observed via broad Lyman-$\alpha$ absorption features in quasar spectra.  They account for the majority of neutral hydrogen out to $z=5$, making them likely candidates as the gas reservoirs for star formation.  Mg~II and other singly ionized metals are also found at redshifts coincident with DLAs -- enrichment that supports the link to star formation.  Past surveys have used Zn and other volatile elements (Si, S, and O), showing that the majority of DLAs have sub-solar metallicities.  The  median metallicity is [Z/Z$_{\odot}$] $\sim -1.2$ \citep{Pett2006}, which increases with decreasing redshift from [Z/H]~$= -2.5$ to $-0.5$ \citep{Wolfe2005}.  DLAs are also likely to contain dust, as the amount of Fe -- a refractory element that condenses at low temperature -- is observed to be depleted relative to Zn and Si with increasing metal abundance \citep{Wolfe2005}.  The extinction profiles for Mg~II and other metal absorbing systems also show significant evidence for reddening by dust \citep{Men2005b,York2006}.

The link between DLAs and dust is strong, but the exact nature of these systems is unknown.  If they are galaxies, observing their emission is difficult because they lie along the line of sight to quasars.  Their sub-solar metallicities also make it difficult to reconcile the idea that DLAs are the progenitors to the galaxies we see today.  They may be dwarf galaxies, inflow, or outflow gas associated with high-$z$ bulges or star forming regions \citep{Wolfe2005}.  X-ray scattering can test these hypotheses because (i) it requires a bright X-ray point source, which the background quasar can provide; (ii) the redshift to the dusty body is known, removing some of the degeneracy in interpreting the scattered image; and (iii) the dust distribution and size of the dusty region places limits on the observed angular size of the scattering halo (\S\ref{sec:ClumpyIGM}).


\section{ X-ray Scattering Cross-Section }
\label{sec:Cross-Section}

The Rayleigh-Gans (RG) approximation is very often applied in the case of X-rays and dust because the index of refraction is close to one and thus the scattering angles are very small \citep[\textsl{e.g.}][]{vdHbook,Over1965}.  The approximation treats a scattering particle as a collection of infinitesimally small Rayleigh scattering regions.  The phase functions for the collection of Rayleigh scatterers must be integrated over the volume of the particle, and relies primarily on its geometry \citep{Martin1970}.  For spherical particles with radius $a$, the angular dependence for differential cross section can be approximated with a Gaussian distribution \citep{MG1986}:
\begin{equation}
\label{eq:dsigma}
\frac{d\sigma }{d\Omega} = \frac{4a^2}{9} \ \left( \frac{2 \pi a}{\lambda} \right)^4 \ |m-1|^2 \ \exp \left( \frac{ -\theta_{\rm scat}^2 }{ 2 \charsig^2  } \right)
\end{equation}
where $m$ is the complex index of refraction and the width of the scattering angle distribution is
\begin{equation}
\label{eq:charsig}
	\charsig = \frac{1.04 \ {\rm arcmin}}{ E_{\rm keV} \ a _{\mum} } .
\end{equation}
Here $a_{\mum} = a / \mum$ and $E_{\rm keV} = E / {\rm keV}$.
The total scattering cross-section is
\begin{equation}
\label{eq:sigma}
	\sigma_{\rm RG} = 2 \pi a^2 \left( \frac{ 2 \pi a }{ \lambda } \right)^2 |m-1|^2
\end{equation}
and is independent of the Gaussian approximation.
We follow \citet{SD1998} in applying the Drude approximation, so that
\begin{equation}
\label{eq:drude}
	| m-1 | \approx \frac{ n_e r_e \lambda^2 }{ 2 \pi }
\end{equation}
where $r_e$ is the classical electron radius and $n_e$ is the number density of electrons, which we calculate from the mass density.
For values characteristic of dust grains,
\begin{equation}
\label{eq:sigmaval}
	\sigma_{\rm RG} = 6.18 \times 10^{-7} \ a_{\mum}^4 \ E_{\rm keV}^{-2} \ \rho_3^2 \ \ {\rm cm}^2
\end{equation}
where $\rho_3 = \rho / 3 \ {\rm g \ cm}^{-3}$, a typical density for dust grains.

\citet{SD1998} have shown that the Rayleigh-Gans approximation matches the more exact Mie scattering solution when $a_{\mum} \lsim E_{\rm keV}$.  For the 0.5 keV photons and 0.25 $\mum$ grains, the magnitude of the scattering cross-section can differ by 20\% (for graphite grains) to 100\% (for silicate grains).  There is also a limiting angle beyond which Equation~\ref{eq:dsigma} no longer matches the exact RG or Mie solution. For 2 keV photons and $0.1 \ \mum$ ($0.25 \ \mum$) grains, the solutions diverge around $500''$ ($200''$).  For soft X-rays $< 1$~keV, the Gaussian approximation matches the shape of the Mie solution out to at least $1000''$.  However, absorption effects are no longer negligible below 1 keV, and the RG approximation causes an overestimate for the halo brightness by a factor of $2-10$ at $\theta_{\rm scat} = 100''$.  This effect also increases dramatically with grain size.  For more details, we refer the reader to the original paper.

Whether or not the $a_{\mum} \lsim E_{\rm keV}$ rule of thumb holds, we can be reasonably confident that Equation~\ref{eq:dsigma} well describes the {\it shape} of the halo image out to $\sim 100''$ and brightnesses for photon energies $> 1$ keV.  A formalism that incorporates Mie scattering is required to extend the accuracy of the calculation to energies $\sim 0.5$ keV.


\section{ X-ray Scattering Through a Uniform IGM }
\label{sec:UniformIGM}

\begin{figure}[t]
\centering
	\includegraphics [scale=0.32] {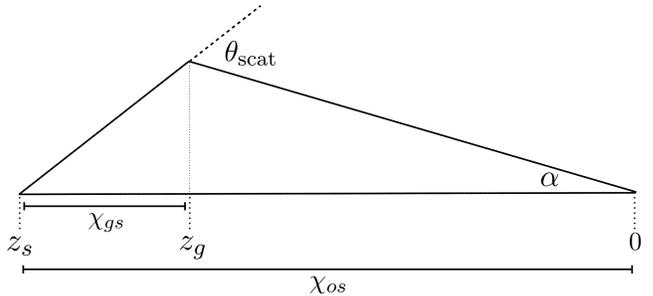}
	\caption{A diagram illustrating the scattering geometry in comoving coordinates.  Photons that scatter through angle $\theta_{\rm scat}$ from a patch of dust grains at redshift $z_g$ will be observed an angular distance $\alpha$ away from the center of the point source. }
	\label{fig:SOdiagram}
\end{figure}
Here we derive the integral for scattered light similar to \citet{Evans1985}, but for an updated $\Lambda$CDM cosmology.
The path taken by light scattered at cosmological distances can be traced over three events:
\begin{enumerate}
\item Light of frequency $\nu_{\rm em}$ is emitted from a point source at redshift $z_s$ and reaches a patch of dust grains at redshift $z_g$.
\item A portion of the incident light is scattered by the dust grain patch. 
\item Scattered light projected onto $\theta_{\rm scat}$ will reach the observer at $z=0$ and be observed an angular distance $\alpha$ away from the center of the X-ray point source.
\end{enumerate}
We use the notation $a_c = (1+z)^{-1}$ to represent the cosmological scale factor throughout.  Everywhere else, $a$ denotes grain radius.

Figure~\ref{fig:SOdiagram} illustrates the geometry of the system using $\chi$ to represent the {\it comoving coordinate distance}:
\begin{equation}
\label{eq:chidef}
	\chi_{12} = \int_{z_1}^{z_2} \frac{c dz}{H(z)}.
\end{equation}
We assume a flat $\Lambda$CDM universe such that $H(z) = H_0 \sqrt{ \Omega_m (1+z)^3 + \Omega_\Lambda }$, where $H_0 = 75$ km/s/Mpc, $\Omega_m = 0.3$, and $\Omega_\Lambda = 0.7$.

In general, the specific flux from a source at cosmological distances is 
\begin{equation}
	F_{\nu} = \frac{L_\nu(\nu_{\rm em}) }{ 4 \pi R^2} \ a_c
\end{equation}
where $R$ is the \textsl{proper radial distance} between the observer and the source.
In the reference frame of a dust grain at $z_g$, the radial distance to the light source is $R_{gs} = \chi_{gs} / (1+z_g)$.  The scale factor of the universe, according to the dust grain, is $a_{c,g} = (1+z_g)/(1+z_s)$.  The flux at the site of the dust grain is therefore
\begin{equation}
\label{eq:Fgrain}
	F_{\nu,g}^{\rm src} = \frac{L_{\nu}^{\rm src} (\nu_{\rm em}) }{ 4 \pi \chi_{gs}^2 } \frac{(1+z_g)^3}{(1+z_s)}.
\end{equation}
We use the subscript notation $\nu,g$ to represents the frequency of light at the dust grain site $[\nu_g = \nu_{em} (1+z_g) / (1+z_s)]$ and $\nu,0$ for the frequency of light at the site of the observer $[\nu_0 = \nu_{em} / (1+z_s)]$.

When light interacts with the patch of dust grains, a small amount will be scattered, with  intensity
\begin{equation}
\label{eq:dIscat}
	dI_{\nu,g}^{\rm scat} = 
	F_{\nu,g}^{\rm src} \ n(z_g) \ 
	\frac{d\sigma_{\nu,g}}{d\Omega} \ dR
\end{equation}
where $n(z_g)$ is the number density of dust grains at the scattering site and $dR$ is the small length
\begin{equation}
\label{eq:dR}
	dR = \frac{ c \ dz }{ H(z) (1+ z_g) } .
\end{equation}
Only light scattered onto the angle $\theta_{\rm scat} = \alpha / x$ will be received by an observer at $z=0$, using the parameterization $x = \chi_{gs} / \chi_{os}$.
Light from the scattering event will reach an observer at $z=0$ with intensity 
\begin{equation}
\label{eq:IobsPerSca}
	dI_{\nu,0} = \frac{dI_{\nu,g}}{(1+z_g)^3}
\end{equation}
Combining Equations \ref{eq:Fgrain} through \ref{eq:IobsPerSca} gives the observed intensity of scattered light from the patch of dust grains at $z_g$:
\begin{equation}
\label{eq:dIscatUni}
	dI_{\nu,0}^{\rm scat} (z_g, \alpha) = 
	F_{\nu,g}^{\rm src} \ 
	\frac{ n (z_g) }{ (1+z_g)^4 } \ 
	\frac{ c \ dz }{ H(z) } \ 
	\frac{d\sigma_{\nu,g}}{d\Omega} \left( \frac{\alpha} {x} \right)
\end{equation}

We can also substitute $n(z) = n_c(z) \ (1+z)^3$, where $n_c(z)$ represents the {\it comoving number density} as a function of redshift.   Drop the subscript for integration over all $z$ and substitute the flux of the source as observed at $z=0$:
\begin{equation}
	F_{\nu,0}^{\rm src} = 
	\frac{ L_\nu^{\rm src} (\nu_{\rm em}) }{ 4 \pi \chi_{os}^2 (1+z_s) }
\end{equation}
to get the total intensity of scattered light as a function of angular distance from the center of a point source.  
\begin{align}
\label{eq:Iobs}
	I_{\nu,0}^{\rm scat} (\alpha) = \ & 
	F_{\nu,0}^{\rm src} \ \times \\
	&
	\int_0^{z_s} n_c (z) \ 
	\frac{(1+z)^2 }{ x^2 } \ 
	\frac{d\sigma_{\nu,z}}{d\Omega} \left( \frac{\alpha} {x} \right) \ 
	\frac{ c \ dz }{ H(z) } \nonumber
\end{align}
The differential cross-section must be evaluated at the frequency encountered by the dust grains [$\nu_z = \nu_0 (1+z)$].

Equation~\ref{eq:Iobs} is missing the effects of attenuation along the path traveled by scattered light.  For a sufficiently homogeneous IGM, the optical depth to extinction along the path of scattered light ($\tau_{gs} + \tau_{og}$) will be nearly identical to that encountered by light traveling in a straight line between the source and observer ($\tau_{os}$).  The extra distance traveled by scattered light, $\Delta \chi \sim10$ kpc (\S\ref{sec:PathDifference}), is not large enough to accumulate an appreciable column density for extinction.\footnote{A notable exception occurs if a galaxy happens to lie along the line of sight.}  Under this assumption, a factor of $e^{-\tau_{os}}$ should modify Equation~\ref{eq:Iobs}.  Then the source flux term can be replaced with the {\sl apparent} flux $F_{\nu,0}^{\rm obs} = F_{\nu,0}^{\rm src} \ e^{-\tau_{os}}$.  Thus throughout the paper we express the halo intensity as a fraction of the central point source apparent brightness.

We use the notation
\begin{equation}
	\label{eq:dfdOmega}
	\frac{ d\Psi_{\nu} (\alpha) }{ d\Omega } = \frac{ I_{\nu,0}^{\rm scat} (\alpha) }{ F_{\nu,0}^{\rm obs} }
\end{equation}
to describe the normalized surface brightness as a function of angle.  
The total optical depth to scattering is
\begin{equation}
	\label{eq:generalSigma}
	\tau_{\rm x} = \int_0^{z_s} \sigma_{\nu,z} \ n_c(z) \ (1+z)^2 \frac{c \ dz}{H(z)}
\end{equation}
for a point-source at $z = z_s$.  Approximately $\tau_{\rm x}^2$ of the source photons will scatter twice, but for the cases considered below, the optical depth to X-ray scattering is always $\tau_{\rm x} \lsim 0.1$.  Therefore we do not need to consider light that might be twice-scattered into or out of the observer's sight.

\subsection{ Dust Grain Distribution }
\label{sec:dust-dist}

We use $\Omega_{\rm dust} = 10^{-5}$ as the fiducial value of comoving IGM dust mass.  The number density of dust grains for identically sized spherical particles is then
\begin{equation}
	n_0 = 8.7 \times 10^{-24} \ h_{75}^2 \ a_{\mum}^{-3} \ \rho_3 \ 
	( \Omega_{\rm dust}^{\rm IGM} / 10^{-5} ) \ 
	{\rm cm}^{-3}
\end{equation}
where $h$ is $H_0$ in units of 75 km/s/Mpc.  The dust we might expect to find in the IGM would have been expelled as a consequence of galactic feedback on the ISM.  We follow the interstellar dust models of \citet[][hereafter MRN]{MRN1977} and \citet[][hereafter WD01]{WD2001} to choose a power law distribution of grain sizes.

The largest grains may be expelled from galaxies more efficiently because they contribute to the opacity of surrounding gas, receive more radiation pressure, and can withstand gas drag and sputtering processes that slow down or destroy small grains (\S\ref{sec:Feedback}).  For the lower end of the size distribution, we choose $0.1 \ \mum$; dust grains smaller than this will be destroyed by $T \sim 10^5 - 10^6$ K halo gas \citep{Ferr1991,Ag1999b}.  We choose an upper limit of $1 \ \mum$ to balance the easy expulsion of large dust grains with the existence, but severe drop in number, of grains $\gsim 1 \ \mum$.  A power law with $p=4$ matches the $R_V = 5.5$ WD01 graphite distribution in the region $0.1 \leq a \leq 1.0 \ \mum$.  To explore the possibility that the dust population resulting from outflow has a larger proportion of $\sim 1 \ \mum$ size grains, we also try a $p=3$ power law throughout the paper.
In the limiting case that all intergalactic dust is large, we will compare results to a dust population solely comprised of $1 \ \mum$ sized grains.

The RG scattering cross-section as used in this study does not incorporate subtleties associated with different grain materials (silicate or graphite).  The complex index of refraction $m$ is calculated from the number density of electrons, which for convenience we assume is uniform regardless of grain material (Equation~\ref{eq:drude}).  Only the size range and power-law slope of the grain distribution affect the X-ray scattering intensities presented throughout this work. \\

\subsection{ The Surface Brightness of Scattered X-ray Light }
\label{sec:uniform-halos}

\begin{table}[t]
\begin{center}
\caption{Optical depth to X-ray scattering for 1 keV photons}
\label{tab:TauCosm}
\begin{tabular}{ c c | c c c c c }
\hline
\multicolumn{2}{c |}{\bf Dust Model}  &  $d\tau_{\rm x} / dR$  &  \multicolumn{4}{c}{ $\tau_{\rm x}$ [\%] } \\ 
{\sl Power}  &  {\sl Sizes} ($\mum$)  &  [\% Gpc$^{-1}$]  &  
  $\mathbf{z = 1}$  &  {\bf 2}  &  {\bf 3}  &  {\bf 4} \\ 
\hline
  --  &  1.0  &  1.68
&  5.2
&  8.1
&  9.9
&  11.2 \\ 
  $(p=3)$  & 0.1 - 1.0 &  0.92
&  2.8
&  4.5
&  5.5
&  6.2 \\ 
  $(p=4)$  &  0.1 - 1.0 &  0.66
&  2.0
&  3.2
&  3.9
&  4.4 \\ 
\hline
\end{tabular}
\end{center}
\end{table}

As an X-ray photon travels between the distant quasar and the observer, its energy will shift towards the softer end of the spectrum.  Thus the total RG scattering cross-section is $\sigma_{\rm RG} \propto a^4 \ E_0^{-2} \ (1+z)^{-2}$, where $E_0$ is the energy of the light observed at $z=0$.  The total optical depth of the universe to X-ray scattering by a uniform, comoving dusty IGM becomes 
\begin{equation}
	\tau_{\rm x} (E_0) = \int_0^{z_s} \frac{c \ dz}{ H(z) } 
	\int_{a_{\rm min}}^{a_{\rm max}}  \sigma_{\rm RG} (a, E_0) \  \frac{dn_c}{da} \ da
\end{equation}
where $a$ is the grain radius.  Table~\ref{tab:TauCosm} shows the optical depth for each dust grain model and $E_0 = 1$~keV.  All of them are significantly below the $\tau_{\rm x} (z=1) \lsim 0.15$ limit set by \citet{DL2009}.  For $p \leq 5$, the optical depth is dominated by the largest grains in the distribution, and the values in Table~\ref{tab:TauCosm} scale roughly as 
$$ \tau_{\rm x}(E_0) \propto \Omega_{\rm dust}^{\rm IGM} \ a_{\rm max}^{5-p} \ E_0^{-2}. $$

The energy dependence of the RG differential cross-section cancels everywhere in Equation~\ref{eq:dsigma} except the Gaussian term.  To absorb the effect of cosmological redshift, we use the parameter
\begin{equation}
	\label{eq:thetaeff}
	\theta_{\rm eff} = \alpha (1+z) / x
\end{equation}
because the Gaussian has a width $\charsig \propto E^{-1}$.  The scattering halo intensity as a function of observation angle (Equation~\ref{eq:Iobs}) becomes
\begin{align}
\label{eq:IobsNc}
	\frac{ d\Psi_{\nu} (\alpha) } { d\Omega }  = &
	\int_0^{z_s} \ \frac{ c \ dz }{ H(z) } \ \frac{(1+z)^2 }{  x^2 } \ \times \\
	&
	 \int_{a_{\rm min}}^{a_{\rm max}} 
	 \frac{dn_c}{da} \
	\frac{d\sigma_{\nu,0} }{d\Omega} 
	\left( \theta_{\rm eff} \right) \ da \nonumber .
\end{align}
For very small optical depths, Equation~\ref{eq:IobsNc} should equal $\tau_{\rm x}$ when integrated over solid angle.

\begin{figure}[t]
	\centering
	\includegraphics [scale=0.5] {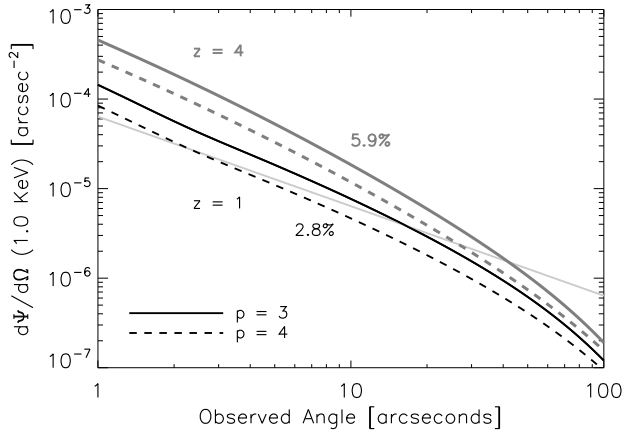}
	\caption{ The normalized surface brightness of scattered X-ray light from a point source at $z=1$ (black) and $z=4$ (grey) using two dust grain distributions: $dn/da\propto~a^{-3}$ (solid line) and $dn/da\propto~a^{-4}$ (dashed line).  The total integrated fractions for the $p=3$ model are shown next to their respective curves. The light grey line illustrates the 100 ks {\sl Chandra} flux limit, normalized by $F_{\rm src}^{\rm obs} = 10^{-11}$ erg/s/cm$^2$ and divided by solid angle. }
	\label{fig:HaloFrac}
\end{figure}

Figure~\ref{fig:HaloFrac} shows the integrated halo profiles for 1 keV X-rays at $z=1$ and $4$.  The total integrated fraction is shown next to each curve for the $p=3$ dust grain model, and in each case agrees with the total optical depth to within a fraction of a percent.  The increased halo intensity that results from using the $p=3$ grain size distribution is not dramatic, around 50\% brighter than the $p=4$ case.  This demonstrates that the halo brightness is more sensitive to the maximum grain size than the power-law exponent.  The light grey line in Figure~\ref{fig:HaloFrac} corresponds to the radial profile that would arise if the brightness within $1''$ spaced annuli summed to the 100 ks {\sl Chandra} flux limit ($4 \times 10^{-15}$ erg/s/cm$^2$), normalized by a point source flux $10^{-11}$ erg/s/cm$^2$.  This illustrates a rough threshold at which it becomes difficult to distinguish signal from background.

The scattering halo becomes more compact with larger source redshift, because the characteristic scattering angle decreases with increasing photon energy (Equation~\ref{eq:charsig}).  For a fixed observed energy band, photons from a high-$z$ source scattered at a higher energy than photons from a low-$z$ source.  Table~\ref{tab:r50} shows the half-light radius (containing 50\% of the total scattered light) for 1 keV X-rays.  Across all grain-size models, scattering halos become more compact by 30\% between redshift 1 and 4.  This creates a dilemma for instruments lacking angular resolution $\sim$  arcseconds.  Point sources at high redshift may have considerably larger column density of dust, but the ability to resolve the resulting scattered halo image will become more difficult.  
\begin{table}[t]
	\caption{Half-light radius for 1.0 keV scattering halo around a source at varying redshift. }
	\label{tab:r50}
\begin{center}
\begin{tabular}{ c r c c c }
\hline
Dust Model & \multicolumn{4}{c}{ Half-light Radius ($''$) } \\ 
  & {\bf z = 1} & {\bf 2} & {\bf 3} & {\bf 4} \\ 
\hline
$1 \ \mum$   & 22 & 18 & 16 & 15 \\ 
$p = 3$   & 31 & 25 & 22 & 20 \\ 
$p = 4$  & 35 & 29 & 26 & 24 \\ 
\hline
\end{tabular}
\end{center}
\end{table}


\section{ X-ray Scattering Through an IGM Clump }
\label{sec:ClumpyIGM}

Though the universe is homogenous on a very large scale, observation along a single sightline is subject to fluctuations in baryon density, as evidenced by the Lyman-$\alpha$ forest and other absorption systems ubiquitous to quasar spectra.  We investigate the possibility that an overdensity in the IGM can be observed individually via X-ray scattering through a screen at intermediate redshift.
As before, Equation~\ref{eq:Fgrain} describes the point source flux from $z_s$ onto a screen of scattering particles at $z_g$.  Equations~\ref{eq:dIscat} and~\ref{eq:IobsPerSca}, implemented with RG scattering (Equations~\ref{eq:dsigma} and~\ref{eq:thetaeff}), produce 
\begin{equation}
\label{eq:IobsScreen}
	 \frac{ d\Psi_{\nu} (\alpha) }{ d\Omega } = 
	\frac{ N_d }{ x^2 } \ 
	\frac{ d\sigma_{\nu,0} }{ d\Omega } \left( \theta_{\rm eff} \right)
\end{equation}
We use the general properties of DLAs as a fiducial example because they are the densest baryon reservoirs observed along the line of sight to known X-ray point sources and are thus the most likely candidates to produce an individual scattering halo.  In addition, we gauge which will contribute more to the brightness and shape of an X-ray scattering halo -- the diffuse IGM or a dense cloud.

The extinction curves measured so far in quasar absorption systems best match the extinction by SMC-like dust.  That is, there is no prominent $2175$~\AA \ feature, but smooth reddening when comparing the spectra of quasars with Mg~II absorbers to those without \citep{York2006,Men2005b,Men2008}.  Extinction curves for the SMC can be well-approximated by an MRN distribution composed purely of silicate grains -- absent are the smallest graphitic grains (PAHs) responsible for the $2175$~\AA \ bump \citep{Pei1992}.  We therefore extend our study to include a distribution of SMC-like dust ($p = 3.5$ and $0.005 \leq a \leq 0.25 \ \mum$), appropriate for star-forming dwarf galaxies.  A distribution of large grains may be more appropriate for dusty regions produced by radiation pressure driven outflows.  As we will show below, the brightness and the width of the X-ray scattering halo offers a diagnostic for differentiating between the two dust populations and their relation to quasar absorption systems.

\begin{table}[t]
\begin{center}
\caption{Potential dust distributions for the average DLA}
\label{tab:DustDist}
\begin{tabular}{l c c | c c}
	\hline
	\multicolumn{3}{c |}{ \bf Dust Model }  &  & \\
	{\sl Name} & {\sl Power} & {\sl Sizes} ($\mum$)  &
	$k$  &
	$\tau_{\rm x}$\footnote{ Optical depth to X-ray scattering at 1 keV for the column density $N_{\rm HI} = 10^{21}$ cm$^{-2}$ and $z=0$.  Note that $\tau_{\rm x} \propto k \ E_0^{-2} \ (1+z_g)^{-2}$. } \\	
	\hline
	SMC-like & $(p=3.5)$  & $0.005-0.25$ &  $3.2 \times 10^{-10}$  &  $\sim 0.5\%$ \\
	Large & $(p=4)$ & $0.1-1.0$ &  $1.7 \times 10^{-13}$  &  $\sim 2\%$ \\
	Large & $(p=3)$ & $0.1-1.0$ &  $6.6 \times 10^{-14}$  &  $\sim 3\%$ \\
	\hline
\end{tabular}
\end{center}
\end{table}

The mass ratio of dust to gas in extragalactic systems may vary with metallicity and size distribution in comparison to that observed locally.  We introduce the dust-to-gas number ratio, $k \equiv N_d / N_{\rm H}$, which absorbs the size distribution, metallicity, and dust-to-gas mass ratio.  If we assume that dust abundance is roughly proportional to metal abundance,
\begin{equation}
\label{eq:Kdef}
	k = \frac{m_{\rm H}}{\langle m_d \rangle} \ 
	\left( \frac{M_d}{M_{\rm H}} \right)_{\rm MW} \ 
	\left( \frac{Z}{Z_{\odot}} \right)	
\end{equation}
where the dust-mass to hydrogen gas-mass ratio of reference is that of the Milky Way, $(M_d/M_{\rm H})_{\rm MW} \approx 0.009$ \citep{DraineBook}.  A relatively constant ratio of dust mass to metallicity is observed locally in studies of the Milky Way, LMC, and SMC \citep{Pei1992}.  We apply this finding liberally as a means of estimating the magnitude of X-ray scattering by dust reservoirs; it may not well apply for dust in outflows, halo gas, or the IGM.  If DLAs are  primordial galaxies, dust detection via X-ray scattering provides a check on dust evolution models \citep[e.g.][]{Inoue2003}.
Table~\ref{tab:DustDist} shows the dust-to-gas number ratio and corresponding optical depth for several dust size distributions, assuming $Z = 1/15 \ Z_{\odot}$.  Note that the total optical depth to X-ray scattering,
$$  \tau_{\rm x} \propto k\ E_0^{-2}\ (1+z_g)^{-2}  $$
will scale proportionally with metallicity and gas-to-dust mass ratio.

\begin{figure}[t]
	\centering
	\includegraphics[scale=0.5]{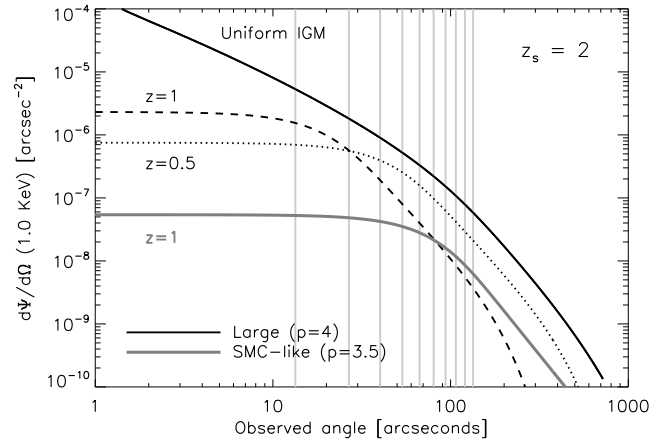}
	\caption{Surface brightness profiles for X-ray photons scattered from a DLA-type screen of dust particles and a point source placed at $z_s = 2$.  The large dust distributions are shown for screens at $z = 1.0$ (dashed black line) and $z = 0.5$ (dotted black line).  For angles $< 10''$, both profiles are flatter than that expected from X-ray scattering through uniformly distributed $\Omega_{\rm dust}^{\rm IGM} = 10^{-5}$ (solid black line).  The SMC-like dust distribution is shown for a screen at $z=1$ (solid grey line).  The light grey vertical lines mark the  radial size ($D_A$) of an object at $z=1$, ranging from 0.1 to 1 Mpc using 100 kpc sized steps.}
	\label{fig:CosmScreen}
\end{figure}

For 1 keV photons and a source at $z_s = 2$, the scattered surface brightness profiles from an infinitely large screen with the hypothesized dust column densities are shown in Figure~\ref{fig:CosmScreen}.  For comparison, the scattering halo expected from dust distributed uniformly throughout the universe is also shown (solid black).  The other black lines show screens at $z=1$ (dashed) and $z=0.5$ (dotted) using the Large $p=4$ dust distribution.  The result for a screen of SMC-like dust is also shown (solid grey).  When the characteristic scattering angle is smaller (e.g.\ with higher energy), the halo profile is more centrally focused.  This explains why the surface brightness profile for the $z=1$ (Large dust) screen, despite having a smaller $\tau_{\rm x}$, appears brighter than the $z=0.5$ (Large dust) screen for observation angles $\lsim 10''$.  When the characteristic scattering angle is larger (e.g.\ with smaller dust grains), the majority of the scattered light is in the wings of the halo profile.  Thus the curve from the SMC-like screen appears significantly dimmer than one might expect in relation to the other curves plotted in Figure~\ref{fig:CosmScreen}.

It is evident from Figure~\ref{fig:CosmScreen} that the expected DLA dust column produces a scattering halo that is several orders of magnitude dimmer than what is expected from $\Omega_{\rm dust}^{\rm IGM} \sim 10^{-5}$, particularly within $10''$ of an X-ray point source.  Therefore, a {\sl typical} DLA along the line of sight to an X-ray bright quasar would not be identifiable with X-ray scattering unless the total dust density $\Omega_{\rm dust}^{\rm IGM} < 10^{-7} - 10^{-6}$.   
However, a high-metallicity DLA (or an extreme dust-to-gas ratio) would present a significant dust column density for scattering.  The rare DLA with $Z \sim Z_{\odot}$ \citep[e.g.][]{Fynbo2011} would have $\tau_{\rm x} \sim 0.08-0.45 \  (1+z_g)^{-2}$, and double-scattering effects will become important.  Eleven DLAs with [Zn/H]~$> -0.5$ (roughly correlating with total metallicity) are present in a sample by \citet{Pro2007}, suggesting $\tau_{\rm x} \gsim 0.02 - 0.14 \ (1+z_g)^{-2}$.

The shape of the halo profile provides a crucial diagnostic for differentiating between a uniformly distributed IGM and a discrete dust source.  
The uniform IGM case is equivalent to stacking many dust-scattering screens between the source and observer.  The more distant screens contribute only to scattering observable at very small angles -- close to the point source -- giving the radial profile a cusp near the center.  For a scattering screen, the halo surface brightness has the same shape as the scattering cross-section (Equation~\ref{eq:IobsScreen}), which we approximated with a  Gaussian function (Equation~\ref{eq:dsigma}).  Therefore, the radial profile of scattered light from a screen is very flat for angles below the characteristic Gaussian width.  The turn-over point, at which the brightness profile falls rapidly, is a function of (i) the dust grain size distribution and (ii) the distance between the X-ray point source and dust screen.  Dust distributions that are weighted towards smaller grains form broader halos because the differential cross-section allows for larger characteristic scattering angles (Equation~\ref{eq:charsig}).  Wider halos are also observed when the light rays approaching the dusty screen are nearly parallel; in comparison, a screen very close to the X-ray point source must scatter light through larger angles in order to reach the observer.  

For a DLA scattering halo to be observable, the dusty region must subtend a large enough angular diameter to be viewed at sizes comparable to the angular width of the scattering cross-section.  
The angular diameter distance, which describes the physical size of an object viewed at cosmological distances,
\begin{equation}
\label{eq:DA}
	D_A = \frac{ \alpha }{ (1+z) } \ \int _0^z \frac{c \ dz}{H(z)}
\end{equation}
is relatively constant beyond $z \sim 1$, for a fixed observation angle.
The light grey vertical lines in Figure~\ref{fig:CosmScreen} mark the angular diameter distance for an object at $z=1$, in 100 kpc steps, from 0.1 to 1 Mpc.  Most objects $z \gsim 1$ must be larger than 200 kpc for the brightness profile turn-off point to be observed, but DLA systems are tens of kpc large, not hundreds \citep{Pett2006}.  Observable X-ray scattering in the vicinity of quasar absorption systems is likely only if these systems are associated with a nearby galaxy, produce outflows or are a tracer of galactic outflow themselves.  If the diffuse distribution of the IGM is sufficiently low, X-ray scattering offers an avenue for exploring the structure around DLAs.


\section{ Redshift Sensitivity }
\label{sec:Zsensitivity}

\begin{figure}[t]
\centering
	\includegraphics[scale=0.5]{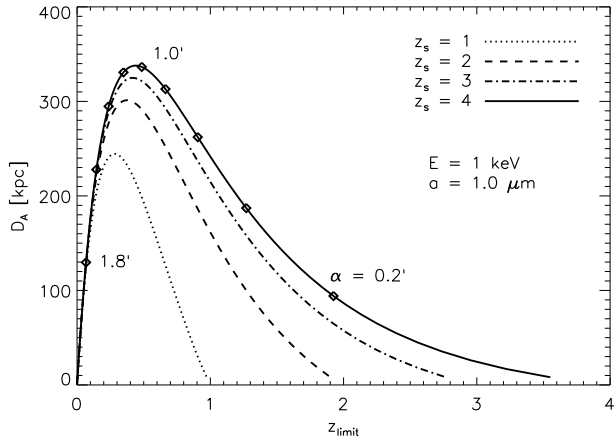}
	\caption{ The physical radius associated with $z_{\rm limit}$, using a fixed photon energy and grain size.  The diamond points mark the observation angle $\alpha$ associated with each $z_{\rm limit}$ in intervals of $0.2'$, the values increasing from right to left. For clarity, $\alpha$ points are only plotted up to $1'$ on the curves below $z_s = 4$.  The results for $\alpha$ and $D_A$ shown here can be scaled by $\sigma_0$(arcmin) $\approx 1 / (a_{\mum} E_{\rm keV})$. }
	\label{fig:zlimit}
\end{figure}

The ability for a region of dust to scatter light towards an observer depends primarily on the scattering angle and the width the Gaussian distribution in Equation~\ref{eq:dsigma}.  We showed with Equation~\ref{eq:thetaeff} that one can imagine all observable scattering is limited to angles $\theta_{\rm eff}$ that fall within a distribution of width $\charsig_0 \equiv \charsig(E_0)$.  Most light will scatter within $\theta_{\rm eff} < 2 \charsig_0$.  Given a particular grain size and energy, for every observation angle $\alpha$ there will be a limiting redshift ($z_{\rm limit}$) beyond which effectively no light will be scattered towards the observer.  Smaller observation angles are able to probe regions closer to the X-ray point source, while larger observation angles are sensitive only to regions close to the observer.   

For 1 keV photons and 1.0 $\mum$ grains, the effective scattering angle is limited by $2 \charsig_0 \approx 2'$.  Figure~\ref{fig:zlimit} shows the angular diameter distance $D_A$ for objects as a function of $z_{\rm limit}$ (found by solving $\theta_{\rm eff} = 2 \charsig_0$ numerically).  The curves border the volume of the IGM that contributes to a scattering halo.  Diamond points on each curve mark the observation angle $\alpha$ in $0.2'$ sized intervals.  To extend the results of Figure~\ref{fig:zlimit} to other grains sizes and photon energies, the $\alpha$ and $D_A$ values should be scaled by $\sigma_0$ in units of arcmin.

The contents of Figure~\ref{fig:zlimit} give broad insight for the ability of X-ray scattering to probe different regions of the IGM.  The contribution to the scattering halo for angles $\alpha > 0.2'$ and a point source at $z_s = 4$ covers a depth $z \lsim 2$ of a uniformly dusty IGM.  An observer who wants to learn more about $z > 2$ dust must be able to measure the halo brightness profile for angles within $\sim 10''$ of the point source center.   The maximum for each curve in Figure~\ref{fig:zlimit} occurs when $\alpha \approx \charsig_0$ ($1'$).  Therefore, the largest detectable extragalactic structures are at $z \sim 0.5$ and have a physical diameter $\sim 500-700$ kpc, depending on the redshift of the background point source.  

When measuring the width of a scattering halo, there is ambiguity between a large object far away and a small object nearby.  From the $z_s=4$ curve in Figure~\ref{fig:zlimit}, a dusty cloud 200 kpc in radius would create a scattering halo of radius $0.2'$ if placed at $z_g=2$.  From Equation~\ref{eq:DA}, a 90 kpc cloud at $z_g = 1$ happens to have an angular radius of $0.2'$ and would thereby create a scattering halo of comparable size.  Knowledge of the object's redshift is required to break the degeneracy between size and distance, making quasars with known absorption systems more favorable candidates for X-ray scattering studies.  Due to the nature of small angle scattering, halo light appearing at $\alpha \approx 2 \charsig_0$ will always have come from dust close to observer.  This fact may help mitigate the effects of foreground dust, which will contribute to the scattering halo several arcminutes from the point source.

\begin{figure*}[t]
\begin{center}
	\subfigure[Changing observation angle]{
		\includegraphics[scale=0.45]{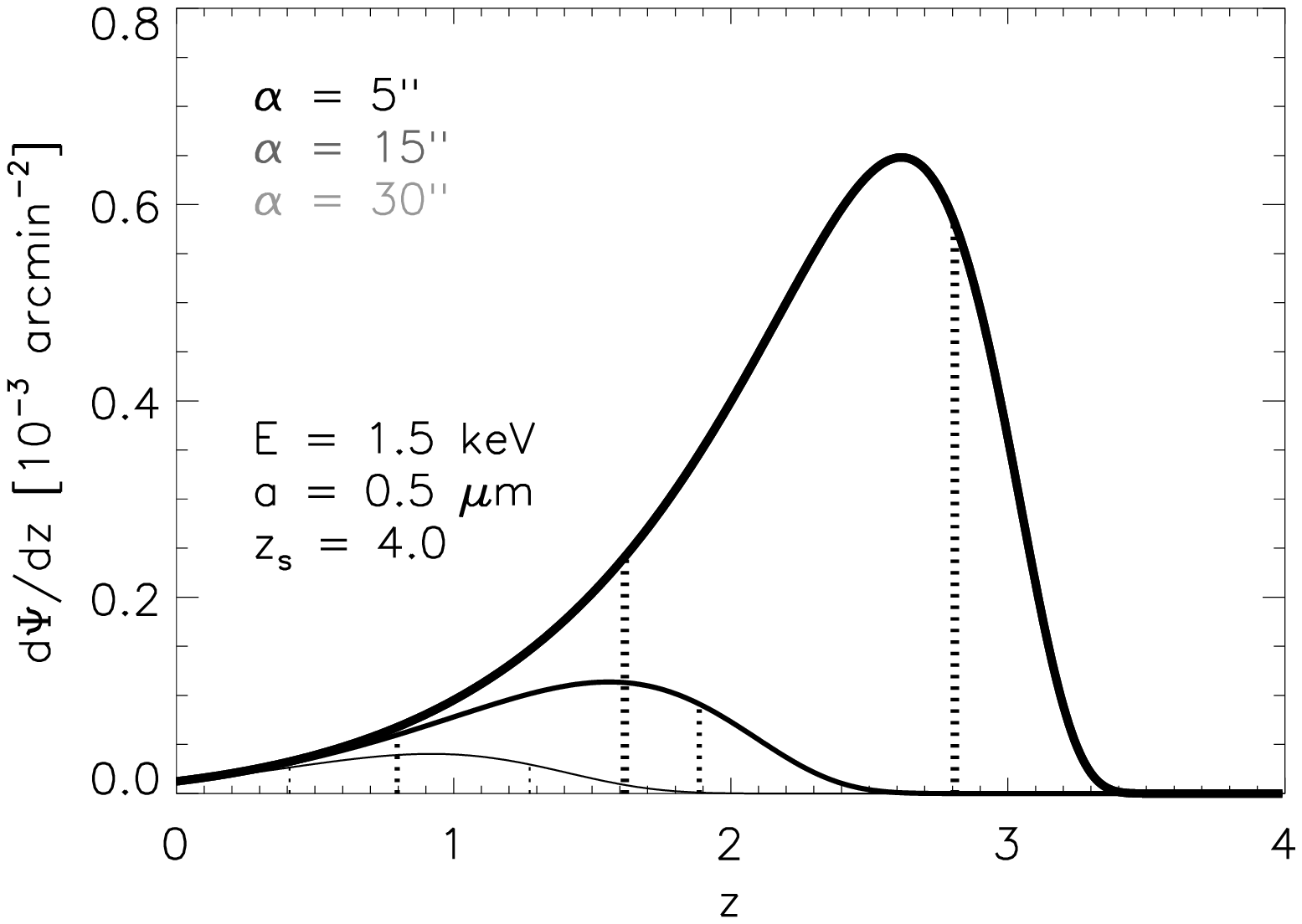}
	}
	\subfigure[Changing photon energy]{
		\includegraphics[scale=0.45]{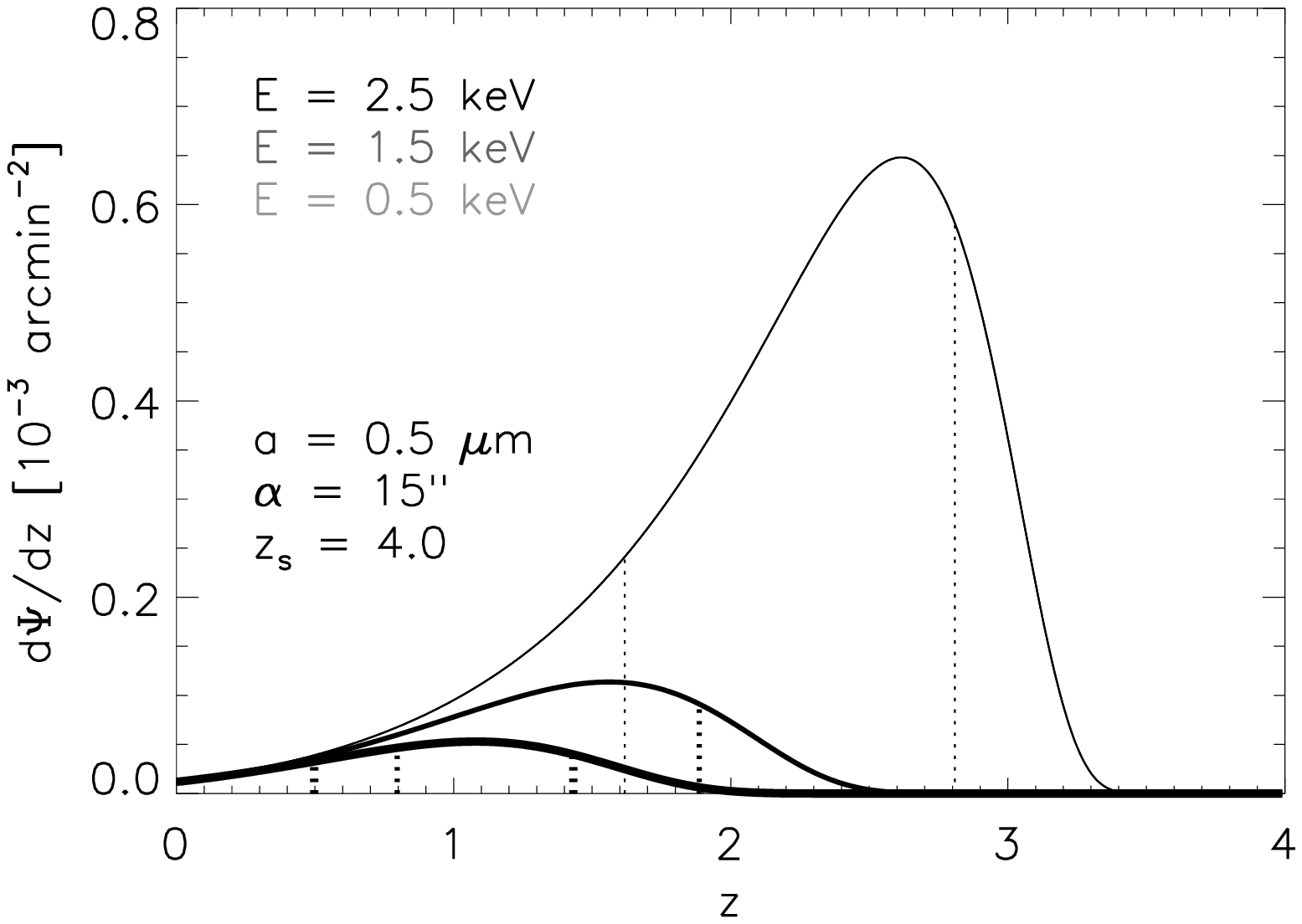}
	}
	\\
	\subfigure[Changing grain size]{
		\includegraphics[scale=0.45]{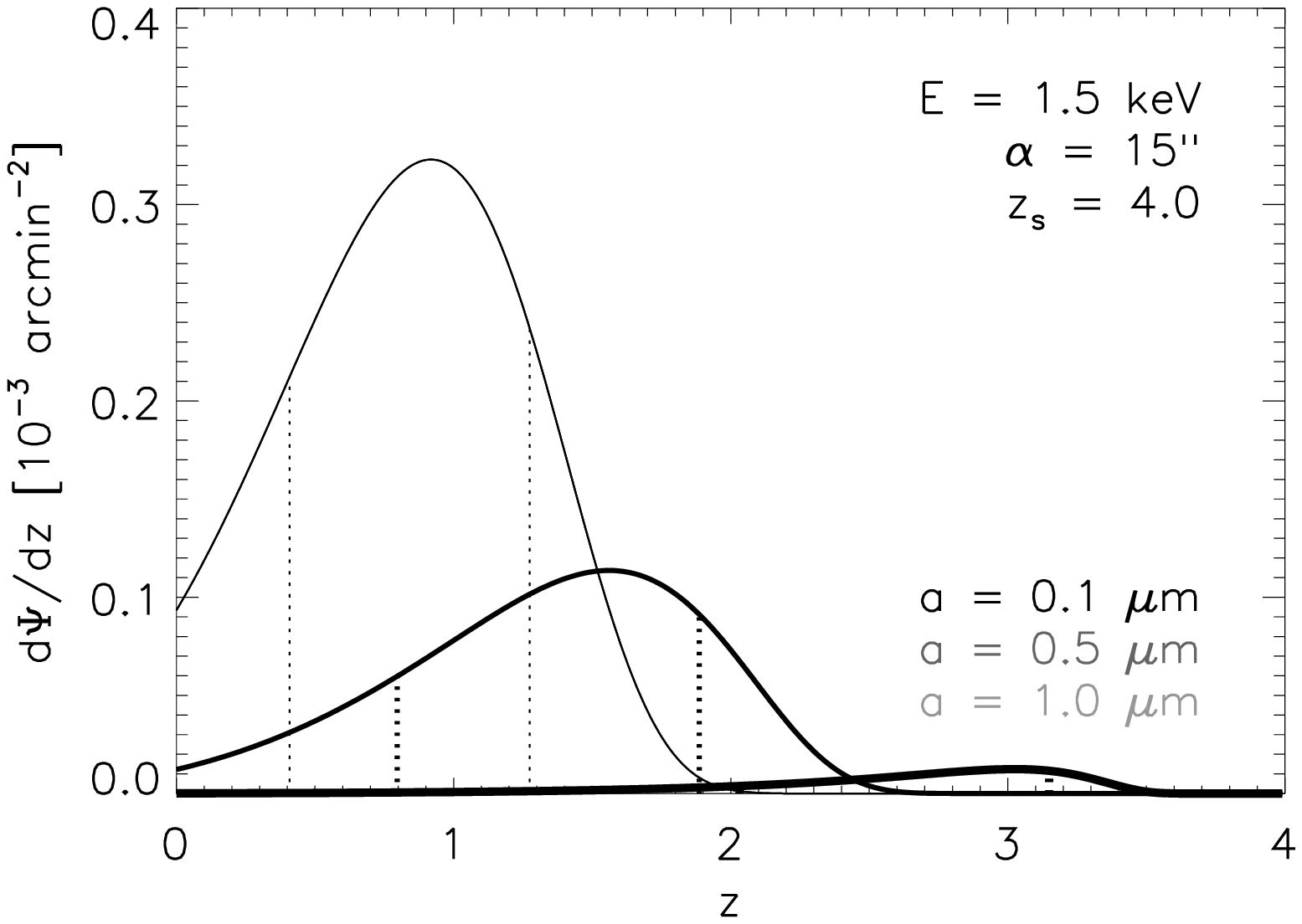}	}
	\subfigure[Changing source redshift]{
		\includegraphics[scale=0.45]{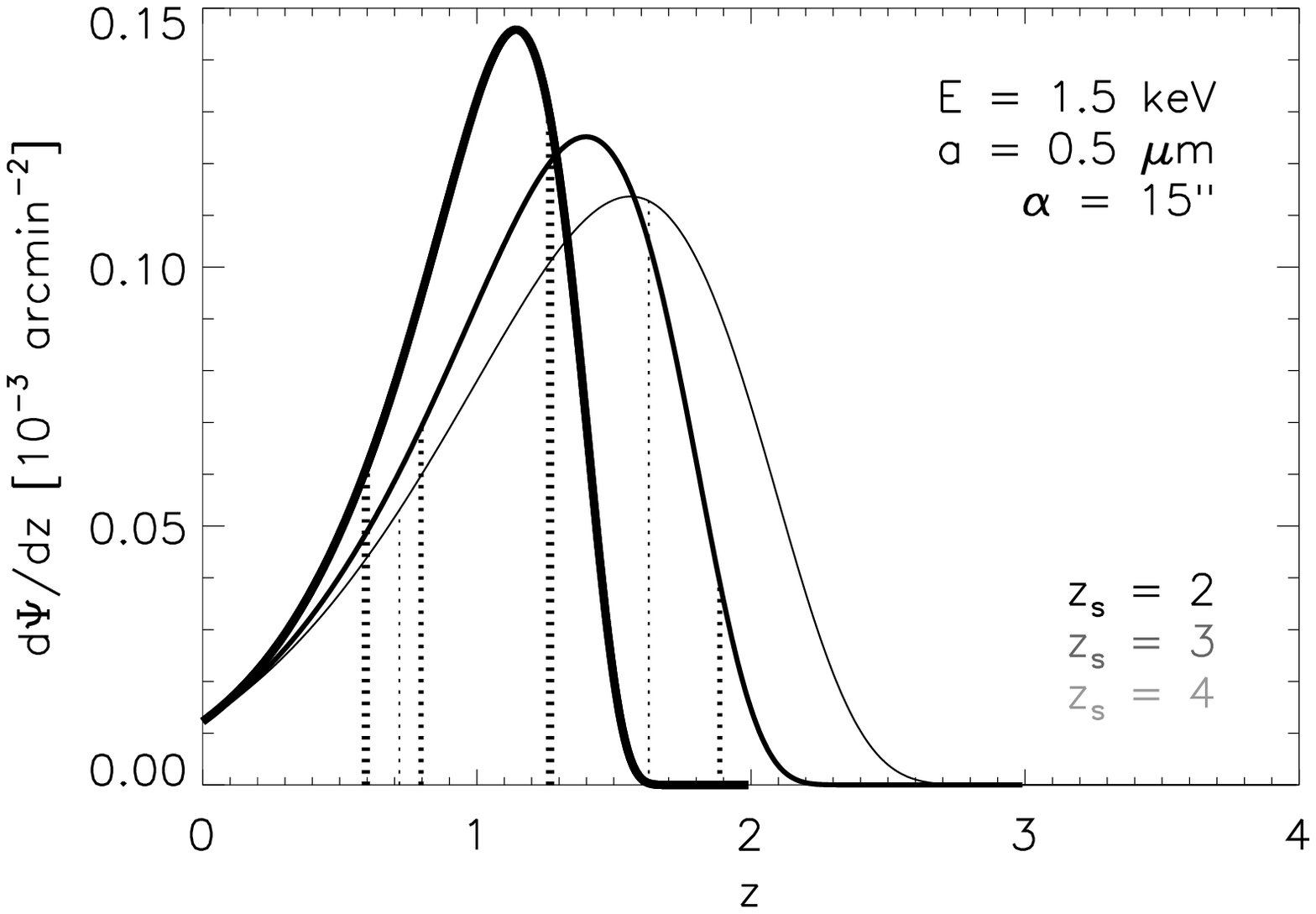}
	}
	\caption{ The curve $\frac{d}{dz} \left[ \frac{d\Psi}{d\Omega} \right]$ as it changes around the case $\alpha = 15''$, $a = 0.5\ \mum$, $E = 1.5$ keV, and $z_s = 4.0$.  The vertical dotted lines mark the central region that encloses 2/3 of the total area under the curve. }
	\label{fig:AlphSens}
\end{center}
\end{figure*}

Each annulus of an X-ray scattering halo image is sensitive to the interplay of several parameters ($a, E,$ and $z$).  We examine the curve
\begin{equation}
\label{eq:dfdz}
	\frac{d}{dz}\left[ \frac{d\Psi}{d\Omega} \right] 
	\propto a^6 \ \Omega_{\rm dust}^{\rm IGM} \ 
	\frac{(1+z)^2}{x^2 \ H(z)} \ 
	\exp\left( \frac{-\theta_{\rm eff}^2}{ 2 \charsig_0^2} \right)
\end{equation}
to gain further insight.
For small values of $x$ (i.e. dust very near the point source), the exponential term becomes very large and the curve asymptotically approaches zero.  The effective scattering angle determines where the curve will turn over and thus which region of redshift which will contribute most to the integral.  Photons that are able to scatter through larger angles will allow sensitivity that extends to higher redshift gas.

Figure~\ref{fig:AlphSens} shows Equation~\ref{eq:dfdz} for different combinations of $\alpha$, $a$, $E$, and $z_s$.  We choose $\alpha = 15''$, $E = 1.5$ keV, $a = 0.5 \ \mum$, and $z_s = 4$ as fiducial values, around which each parameter was varied.  The curves reach a maximum at intermediate redshift, but are often asymmetric.  The vertical dotted lines demarcate the central region for which the area under the curve equals 2/3 of the total integrated brightness at the respective observation angle.

Light that scatters close to the point source must generally go through large scattering angles to be observed.  Therefore, scattered light observed at small angular distances from the point source center predominantly traces gas closer to the source (Figure~\ref{fig:AlphSens}a).  
For the same reason -- low photon energies, which can scatter through larger angles, will be able to probe gas closer to the point-source (Figure~\ref{fig:AlphSens}b).  The curves in Figure~\ref{fig:AlphSens}b are degenerate with Figure \ref{fig:AlphSens}a because the width of the Gaussian, $\charsig \propto \alpha / E$.
Large dust grains, which are more strongly forward scattering than small dust grains, will only contribute observable scattering from regions of low redshift.  This is because gas close to the observer receives light rays that are nearly parallel to the line between the observer and point-source, making the required scattering angle for observation lower compared to dust at intermediate distances (Figure~\ref{fig:AlphSens}c).
Finally, viewing a source at smaller redshift will probe gas closer the observer.  However, the shift in region sensitivity is small; scattering halos around objects at $z = 2-4$ are mostly sensitive to dust in the $z = 0.5 - 2$ range (Figure~\ref{fig:AlphSens}d).
It is interesting that reducing the point source's redshift ($z_s$) does not dramatically change the redshift range to which the halo image is sensitive.  A campaign that uses image stacking to search for an X-ray scattering halo could therefore use broad bins in $z$.

\subsection{Constraints on IGM homogeneity}

Figure~\ref{fig:zlimit} can also be used to place constraints on the homogeneity of the IGM.  If a scattering halo is asymmetrical on an annulus $\sim 1$ arcminute in size, the dust distribution must be inhomogeneous on $\sim 300$ kpc scale.  A symmetric scattering halo would thus indicate a population of dust grains that is smoothly distributed.  Observation angles around $10''$ correspond to spatial homogeneity $\sim 100-200$ kpc in diameter.  A scattering halo that appears asymmetrical across a $10''$ annulus would thus indicate that the dust being probed is confined to the gaseous regions of a galactic halo, potentially part of an outflow.

To illustrate, galactic winds $\sim 100$ km/s would travel about 100 kpc per Gyr.  As a first approximation (and upper limit), the distance traveled by a dust particle that has escaped the potential well of its parent galaxy (at $z_2$) and is thus not subject to any appreciable force of gravity, is
\begin{equation}
	\int_{z_1}^{z_2} \frac{v \ dz}{(1+z) \ H(z)}
\end{equation}
where $v$ is the terminal velocity of the particle.  Assuming $v \sim 100$ km/s for particles that have been ejected from a small galaxy at $z=2$, a dust grain would travel $\sim 200$ kpc by $z=1$.  

Note that if a 200 kpc thick shell of ejected dust continued to expand freely, it would be $\sim 1$ Mpc in radius by $z=0$ and about 75 times less dense than it was was at $z=1$.  A comparable amount of dust grains uniformly distributed throughout the IGM would become less dense by a factor of 8 between $z = 1$ and $z = 0$.  This means that if a uniformly distributed population of intergalactic dust is observed, it is more likely to have been distributed during the epoch $z \gsim 3$ than from $z<2$ galaxy outflows, which dominate the region $R \lsim 100$ kpc around them.

\subsection{Constraints on quasar variability}
\label{sec:PathDifference}

Scattered photons must travel a slightly longer distance than non-scattered photons before reaching the observer.  For small angles, this path difference: 
\begin{equation}
\label{eq:DeltaChi}
	\Delta \chi \approx \chi_{os} \ \alpha^2 \ (1 - x)
\end{equation} 
requires knowledge about the distance to the dust grain.  We showed that below the value $x = \alpha / 2 \charsig$, almost no light will be scattered into the observer's line of sight.  Substituting this into Equation~\ref{eq:DeltaChi} and recognizing that the {\it maximum} path difference occurs for observation angles $ \sim \charsig$, we get
\begin{equation}
	\Delta \chi_{\rm max} \sim 4.5 {\rm \ kpc} \ 
	\left( \frac{ \chi_{os} }{ {\rm Gpc} } \right) 
	\left(\frac{ E }{ {\rm keV} }\right)^{-2} \left(\frac{ a }{ 0.1 \ \mum }\right)^{-2}
\end{equation}
for choices of $E$ and $a$ that give large $\charsig$.

The extra distance traveled by a scattered photon leads to a time delay $ \Delta t \sim 10^4$ years, which is much shorter than the suggested quasar lifetime, $\sim 10^6 - 10^8$ years \citep{Martini2004}.  It is therefore unlikely for an X-ray scattering halo to be truncated due to a particular quasar's activity switching ``on'' recently.  However, dramatic variations in a quasar's brightness on $\leq 10^4$ year timescales would become apparent if a scattering halo exhibit radial variation.  If we traced the region of observable scattering from a short burst of X-ray light, it would take the shape of an elliptical shell.  Thus a ring of increased (or decreased) surface brightness might appear around quasar, offering a means to constrain the long-term variability of AGN activity \citep[e.g.][]{ME1999}.


\section{Comparison to Petric et al. (2006)}
\label{sec:QSO1508}

High spatial angular resolution coupled with a focusing power that confines $\sim 99\%$ of point-source light to a $2''$ (4 pixel) radius makes {\sl Chandra} the best X-ray observatory in operation today for observing scattering halos.  The relatively narrow point-source function (PSF) assures that there is enough contrast between the source and background to observe an accumulation of dust-scattered photons a small angular distance away.  As illustrated below, the most difficult part about observing scattering halos due to intergalactic dust is obtaining a statistically significant number of photon counts with respect to the models.  

\begin{figure}[t]
\centering
	\includegraphics[scale=0.5]{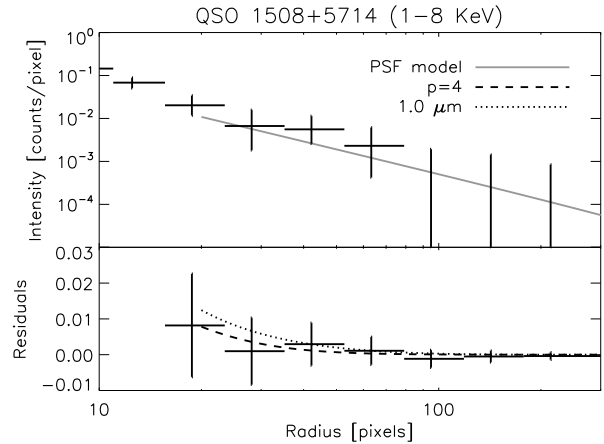}
	\caption{({\it Top}:) The radial profile for QSO 1508+5714 in the 1-8 keV range, with the model point-spread function in grey.\\  ({\it Bottom}:) The residual intensity after subtracting the PSF model of \citet{Gaetz2004}, compared to the intensity expected from a $p=4$ power-law distribution of grey dust (dashed line) and $1 \ \mum$ size grains (dotted line) with total mass density $\Omega_{\rm dust}^{\rm IGM} = 10^{-5}$. }
	\label{fig:qso1508}
\end{figure}

\begin{table}[b]
\begin{center}
\caption{Observed and expected properties for QSO 1508+5714 and its scattering halo.}
\label{tab:QSO1508}
	\begin{tabular}{c c c c}
	\hline
	{\bf Energy (keV)} & 
	$N^{\rm src}_{\rm obs} (< 2'') $\footnote{Number of counts observed within a $2''$ radius of point source center.} & 
	$N^{\rm sca}_{\rm exp} (10-500'')$\footnote{The total number of scattered counts expected to be observed in the region $10''-500''$ away from the point source, rounded to the nearest integer.} &
	$N_{\rm bkg}$\footnote{The estimated number of background counts, extracted from a circular region on the same image, for an area covered by an annulus $10-500''$ wide.} \\ 
	\hline
	0.3 - 1.0 & 1943 & 375 (38)\footnote{With a factor of 10 reduction due to absorption.} & 24809 \\ 
	1.0 - 2.0 & 1915 & 32 & 15074 \\
	2.0 - 3.0 & 637 & 3 & 12247 \\ 
	3.0 - 4.0 & 368 & 1 & 8479 \\
	\hline
	\end{tabular}	
\end{center}
\end{table}

\citet{Petric2006} used the non-detection of an X-ray halo image around QSO 1508+5714, a bright quasar $\sim10^{-13}$~erg/s/cm$^2$ at $z = 4.3$, to place the limit $\Omega_{\rm dust}^{\rm IGM} < 2 \times 10^{-6}$.  They did this by comparing the expected halo brightness from a dust population of $1.0 \ \mum$ size grains to the brightness of the PSF wings predicted by MARX (a ray-trace simulation) and the wings from a previous {\it Chandra} observation of 3C 273.
The number of scattered photons expected from $1.0 \ \mum$ dust is several times that expected from a distribution of dust grain sizes, allowing us to relax their constraints.

The reliability of MARX to simulate the {\it Chandra} PSF wings is uncertain.  High energy X-rays are more subject to deviations in scattering by micro-roughness in the high resolution mirror assembly (HRMA) and require smaller grazing incidence angles that are difficult to obtain while testing the HRMA on the ground.  In addition, appropriately determining the relative brightness of the PSF wings from objects like 3C 273 is difficult due to pile-up -- that is, the tendency of the {\it Chandra} CCD instrumentation to register an incorrect photon energy or mistake multiple photon events for a cosmic ray -- which occurs for bright objects ($\gsim 1$ X-ray photon per second).
Pile-up causes an obvious depression of counts in the center of 3C 273, prevents the accurate measurement of source flux, and thereby degrades the accuracy of the radial profile within $\sim 10''$, or 20 pixels \citep{Gaetz2004}.  Since the radial profile from 3C 273 appears to match QSO 1508+5714 best around $5''$ (10 pixels), the extrapolated wing profile is more likely to be an overestimate of the true PSF -- perhaps explaining why the intensity from 3C 273 matches the upper limits on the instrumental measurement between $30 - 60''$ \citep[][Fig. 2]{Petric2006}.

We use the PSF calibration model  by \citet{Gaetz2004} to gauge the PSF + halo profile that would have resulted from the $p=4$ dust distribution with density $\Omega_{\rm dust}^{\rm IGM} = 10^{-5}$.  A quick examination of QSO 1508+5714 (obsid 2241) using a circular region with a 4 pixel radius yields approximately 5400 counts in the 0.3-8 keV range.  (This is a lower limit to the total number of source counts, as it does not consider counts spread into the PSF.)  For several energy bins, Table~\ref{tab:QSO1508} shows the number of point source counts, expected number of halo counts in an annulus $10-500''$ centered on the point source, and the number of background counts expected in that same region.  The majority of the scattered light will have photon energies $< 2$ keV.  However, for soft X-rays ($< 1$ keV) and large dust grains, the RG cross-section can overestimate the intensity of scattering by a factor $\sim 10$ due to absorption by grain material \citep[][]{SD1998}.  For this reason, the 375 scattered counts expected in the $10-500''$ annular region are likely too large by an order of magnitude.  We therefore exclude photon energies below 1 keV from further evaluation.  To take advantage of the soft X-ray band, the halo profile must be modeled with Mie scattering.

Figure~\ref{fig:qso1508} compares the radial brightness profile of QSO 1508+5714 in the 1-8 keV energy band to the PSF and dust models. The magnitude of the error bars show that the observations of \citet{Petric2006} are consistent with $\Omega_{\rm dust}^{\rm IGM} \lsim 10^{-5}$ for the power-law distribution of dust grains chosen.  However, the brightness profile is also consistent with a population of IGM dust composed purely of $1 \ \mum$ sized grains.  By including energies below 1 keV and using the RG cross-section, \citet{Petric2006} were expecting a much brighter X-ray halo.  This was an overestimate leading to the limit $\Omega_{\rm dust}^{\rm IGM} < 2 \times 10^{-6}$, which our analysis shows can be relaxed to $\Omega_{\rm dust}^{\rm IGM} \leq 10^{-5}$.  This emphasizes the need for a more detailed modeling, incorporating statistics and Mie scattering, to be implemented in the future.  An even more important task would be the careful calibration of the {\it Chandra} HRMA PSF between 1 and $20''$.


\section{Discussion}
\label{sec:Discussion}

X-ray scattering provides a unique view of the cosmic history of star formation, feedback, and the IGM that is complementary to UV, optical, and infrared studies.  A summary of our main results are as follows: \\
\indent (i) If dust is evenly distributed throughout the IGM, with $\Omega_{\rm dust}^{\rm IGM} = 10^{-5}$, the dust-scattered X-ray light will be $\sim 5\% \ (E_{\rm keV}^{-2}) $ of the point source brightness for objects at $z \geq 2$.  The intensity of the scattered light is directly proportional to $\Omega_{\rm dust}^{\rm IGM}$.  \\
\indent (ii) Also in the uniform IGM case, point sources at larger values of $z$ trade larger dust column densities for a narrower halo profile, making it slightly more difficult to distinguish scattered light from the point source. \\
\indent (iii) For high-$z$ point sources, scattered light comes predominantly from intergalactic dust at $z \lsim 2$.  If the majority of IGM dust came from star formation within sub-L* galaxies at $z > 3$, assuming a constant comoving density of dust grains may still be reasonable for point sources at $z = 4$. \\
\indent (iv) If X-ray light scatters through a dense, dusty clump in intergalactic space, the resulting halo image will have a flatter radial profile compared to the uniformly distributed case.  The further away the clump is from the background point-source, the larger the halo image will appear to an observer. \\
\indent (v) If the dust mass to metallicity ratio is relatively constant, as observed locally, then a typical DLA will have an optical depth to 1 keV X-ray scattering $\sim 1-3\%$, proportional to $(1+z_g)^{-2}$. This result scales proportionally with metallicity, so DLAs with near-solar abundances may have $\tau_{\rm x} \gsim 5 \%$. \\
\indent (vi) Dusty clumps at $z \gsim 1$ would have to be $\geq 200$ kpc in size for a halo image to appear at angles larger than $10''$ from the point source center.  X-ray scattering thus offers the opportunity to test if dusty outflows exist near quasar absorption systems, but the outflow may require  column densities of dust that are so far unprecedented.  Perhaps an X-ray point source whose line of sight is close to a foreground galaxy can test the ability of large grains to be expelled from L* galaxies. \\
\indent (vii) A scattering halo that is symmetric on an annulus of radius $\geq 60''$, would indicate that IGM dust is distributed uniformly.  A scattering halo that is  asymmetric on an annulus of radius $10''$ would indicate the the dust distribution is traced mainly by galaxies ($R \lsim 50$ kpc) and is not distributed very far out in the galactic halo potential or the IGM.  Either result would provide clues to galactic and star formation feedback mechanisms.\\
\indent (viii) The previous constraint of $\Omega_{\rm dust}^{\rm IGM} < 2 \times 10^{-6}$ must be relaxed when considering the absorption of X-ray light below 1 keV.  Photons observed at 0.5 keV today will be shifted above 1 keV for $z \gsim 1$, but in Section~$\ref{sec:Zsensitivity}$ we showed that the scattering halo is more sensitive to material at $z \lsim 2$ and large dust grains at low redshift.   Therefore, what would be the dominant contribution to the scattering calculation, X-rays $< 1$ keV and $1 \ \mum$ sized grains, is overestimated by Rayleigh-Gans scattering.  Deeper observation of high redshift quasars, in conjunction with a more careful calculation involving the Mie scattering solution, is required to place a stronger constraint on the intergalactic dust distribution.

\begin{table}[t]
\begin{center}
	\caption{Systematic offset in magnitude for an intergalactic population of grey graphitic grains.}
	\label{tab:opt-ext}
\begin{tabular}{ c | c c | c c }
\hline
  & \multicolumn{2}{| c |}{ $\mathbf{(p=3)}$ } & \multicolumn{2}{| c }{ $\mathbf{(p=4)}$ } \\ 
$\mathbf{z}$  &  $A_V$  & $A_{1.78}$  &  $A_V$  &  $A_{1.78}$ \\ 
\hline
0.50  &  0.008  &  0.007  &  0.012  &  0.010 \\ 
1.00  &  0.010  &  0.010  &  0.017  &  0.014 \\ 
2.00  &  0.012  &  0.012  &  0.019  &  0.017 \\ 
4.00  &  0.013  &  0.013  &  0.021  &  0.019 \\ 
\hline
\end{tabular}
\end{center}
\end{table}

To conclude, we briefly note that the dust models used throughout this paper would create meaningful systematic errors for both optical and infrared extragalactic surveys.  Intergalactic dust dims the most distant objects, and if not accounted for, would affect cosmologists' ability to place more stringent constraints on dark energy models.  A systematic shift as low as $\delta m \sim 0.02$ would reduce the effectiveness of future supernovae surveys to constrain the dark energy equation of state parameter $w$ \citep{Vir2007}.  A systematic shift in magnitude roughly correlates as $\delta m \sim -0.5 \ \delta w$ and $\delta m \sim - \ \delta \Omega_M$ for $z \lsim 1$ \citep[e.g.][]{ZhangP2008, Men2010b, Cora2006}.  
Thus an extinction offset $\sim 1\%$ would create an offset comparable to the error bars sought after in precision cosmology.

A distribution of graphitic grains larger than $0.1 \ \mum$ is grey even to the reddest filters ($0.893 \ \mum$ and $1.78 \ \mum$) to be used in cosmological supernovae surveys (LSST and WFIRST, respectively).  We calculate the systematic offset in magnitude for our grey graphitic grain models using $\Omega_{\rm dust}^{\rm IGM} = 10^{-5}$ (Table~\ref{tab:opt-ext}).There is little offset between $A_V$ and $A_{1.78}$ -- the extinction at $1.78 \ \mum$ -- which is the center of the reddest filter on the WFIRST mission \citep{WFIRST}.
For supernovae surveys beyond $z >1$, $\delta m \sim 0.01 - 0.02$ in both the visual and infrared bands.  Such small systematic offsets would call for a deep field survey, rather than wide field survey that increases the precision of the photometric measurements but not their accuracy \citep{Vir2007}.  Observational searches for X-ray scattering halos due to intergalactic dust would constrain these systematic offsets and help determine the best future course of action for large-scale optical and infrared cosmological surveys.

Finally, we note that the values in Table~\ref{tab:opt-ext} differ by those calculated by \citet{Ag1999b} for a similar cosmic density of intergalactic dust.  For the model $p=4$ graphite grain distribution, $\kappa_V = 2.4 \times 10^4$ cm$^2$/g.  Using this value with $\Omega_{\rm dust}^{\rm IGM} = 10^{-5}$ in Equation~9 of \citet{Ag1999b} implies $A_V (z=0.5) \sim 0.02$.  However, Aguirre uses $\Omega_m = 0.2$ and $\Omega_\Lambda = 0$.  In that case, the distance (and hence the column density) one would calculate for $z=0.5$ is larger by a factor of two than that in the currently accepted $\Lambda$CDM universe.  This brings previous work into agreement with the results shown in Table~\ref{tab:opt-ext}. \\


We thank the anonymous referee for their attention to detail and help in greatly improving the clarity of the paper.  We thank J. G. Peek and J. Schroeder for useful discussion about extragalactic dust and cosmology.  This work was supported in part by NASA Headquarters under the NASA Earth and Space Science Fellowship Program - Grant NNX11AO09H.

\bibliography{references}

\end{document}